\documentclass[pra,twocolumn,10pt,longbibliography]{revtex4-1}

\usepackage{amsmath,stmaryrd}
\usepackage{microtype}

\usepackage{graphicx}
\usepackage{float}
\usepackage[caption=false]{subfig}
\usepackage[export]{adjustbox}

\usepackage[svgnames]{xcolor}
\usepackage{hyperref}
\hypersetup{colorlinks,linkcolor={DarkBlue},citecolor={DarkBlue},urlcolor={DarkBlue}}
\usepackage[capitalize]{cleveref}

\begin{document}

\title{Circumventing defective components in linear optical interferometers}

\author{Ish Dhand}
\affiliation{Xanadu Quantum Technologies, 777 Bay Street,Toronto ON, M5G 2C8, Canada,}

\date{\today}
\begin{abstract}
A crucial challenge to the scaling up of linear optical interferometers is the presence of defective optical components resulting from inevitable imperfections in fabrication and packaging.
This work presents a method for circumventing such defective components including lossy modes and unresponsive phase shifters and beam-splitters.
The method allows for using universal linear optical interferometers with such defects as smaller universal interferometers.
The method presented here tolerates remarkably high defect rates in constructing linear optical interferometers, thus bringing closer to reality the possibility of obtaining quantum advantage with linear optics.
\end{abstract}
\maketitle

\section{Introduction}

Linear optics is a promising route to obtaining a quantum advantage in simulation~\cite{Huh2015a}, computation~\cite{Aaronson2013a,Hamilton2017a,Rudolph2017a}, communication~\cite{Duan2001a} and metrology~\cite{Motes2015a,Olson2017a}.
Furthermore, linear optics also enables novel applications in the domain of optical neural networks and neuro-morphic photonics~\cite{Shen2017,Tait2017,Steinbrecher2019}.
Scaling up linear optical interferometers to a large number of optical modes is critical for obtaining an advantage in these applications.
For instance, provably beating classical computers in a boson sampling task could require interferometers acting on several hundreds of optical modes~\cite{Neville2017,Clifford2018}.
Obtaining an advantage in Gaussian boson sampling~\cite{Hamilton2017a,Kruse2019}, which has many promising applications~\cite{Bradler2018,Arrazola2018,Arrazola2018a,Banchi2019,Jahangiri2019}, could require upwards of tens to hundreds of modes~\cite{Gupt2018,Bjorklund2018,Quesada2019,Wu2019}.
Scaling to more modes requires interferometers comprising a large number of optical components, with the number of beam-splitters and phase-shifters growing as the square of the number of modes.

As the number of components rises, defects arising in optical components during the fabrication and packaging of interferometers become more common.
If an interferometer becomes unusable in the presence of a single defect, then these defects become a key impediment in the scaling up of interferometers to a large number of optical modes.
Currently, this is indeed the case.
In current interferometers that act on a low tens of optical modes (see Ref.~\cite{Harris2018,Flamini2019} for reviews), the usual method of addressing defective components is to fabricate and package multiple interferometers, discarding the defective ones, and choosing one that is free from defects.
This strategy is infeasible for future applications, wherein hundreds to thousands of modes could be required for obtaining and exploiting a quantum advantage.
Such large interferometers would require two to four orders of magnitude more optical components than current interferometers, which leads to a concomitant two-to-four order of magnitude increase in the probability of at least one defect arising in the interferometer.

Integrated linear optical interferometers, which combine multiple optical components on a single chip, are the most promising route to obtaining a quantum advantage with linear optics~\cite{Flamini2019} so we focus on these interferometers here.
An integrated interferometer with a large number of optical components will unavoidably have defects, such as excessively lossy modes and unresponsive beam-splitters.
Lossy optical modes (with up to 10dB excess loss) could result from dust present in the cleanrooms used for fabricating integrated photonic interferometers as has been observed in demonstrations of integrated linear interferometers such as that of Ref.~\cite{Wang2018a}.
A defect of this form, if present in a Mach-Zehnder interferometer (MZI) that is used to realize a beam-splitter, could also make the entire resulting beam-splitter lossy.
These losses render the interferometer transformations non-unitary into a form that is difficult to analyze and exploit.

Another important class of defects in a reconfigurable linear optical interferometer is an unresponsive or partially responsive optical component such as an MZI or phase-shifter.
A component could become unresponsive from imperfections in the fabrication of the chip or because of an imperfect packaging, which could lead to an imperfect interface between the chip and its classical control and processing unit~\cite{Harris2014a}.
Another issue is that of an MZI being only partially configurable, i.e., one whose extinction ratio does not range from $0:100$ to $100:0$; but can only achieve a subset of this range, for instance the range between $5:95$ and $80:20$.
These imperfect ranges often arise in the fabrication of MZIs as any deviation of the composing beam-splitters from exactly $50:50$ will lead to an imperfect extinction-ratio range of the MZI.
Although self-reconfiguring designs to implement ideal MZIs have been proposed~\cite{Miller2015a,Wilkes2016a}, these designs only allow for ideal MZI operation in a limited range of extinction ratios ($15:85$ to $85:15$) and moreover these increase the spatial footprint of MZIs by at least a factor of two.

Unresponsive or partially responsive components would preclude important applications of reconfigurable interferometers including the possibility of implementing arbitrary unitary matrices drawn from the Haar measure as required by boson sampling and Gaussian boson sampling~\cite{Aaronson2013a,Hamilton2017a}. 
Furthermore, these defects would prevent the use of reconfigurable interferometers in a hybrid spatial-temporal setting for implementing large-scale unitary transformations~\cite{Dhand2015,Su2019a}.
Thus, a method that enables using interferometers despite the presence of defective components of all kinds is vital for future applications of linear optics.

The current work provides such a method, which allows for using universal linear interferometers despite defective components.
The method enables circumventing defects in the sense that no light ever reaches these defects whereas the remaining optical components of the interferometer can be used to implement an effectively smaller universal interferometer.
Thus, any interferometer that is fabricated as a reconfigurable universal linear optical interferometer can still be used as a universal interferometer, albeit one acting on fewer modes.

\section{Procedure for circumventing defective optical components}

This section presents the procedure for circumventing defective components. 
The circumvention of a single defect is described in \cref{Sec:Single}.
Next, \cref{Sec:Multi} builds upon the single-mode case in order to circumvent multiple single-mode defects.
This procedure for dealing with multiple single-mode defects also allows for circumventing defects in two-mode optical elements including beam-splitters as a special case.
\cref{Sec:Shallow} presents details on the mitigation of defects in shallow optical circuits.

\subsection{Circumventing single-mode defects}
\label{Sec:Single}
\begin{figure*}
\centering
\subfloat{\includegraphics[scale = 0.85, valign = c]{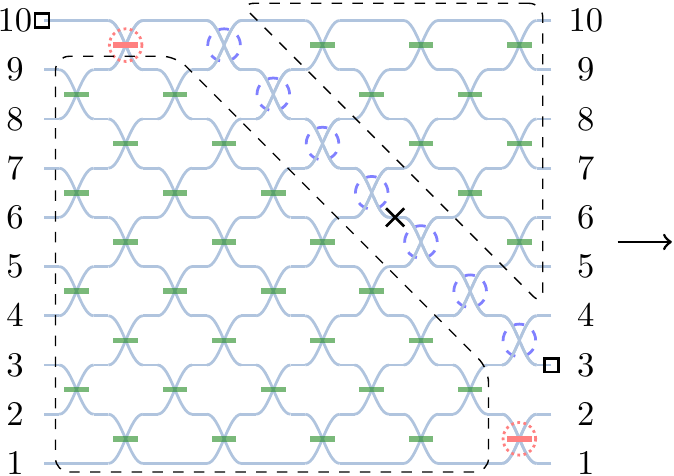}}\hspace{1mm}
\subfloat{\includegraphics[scale = 0.85, valign = c]{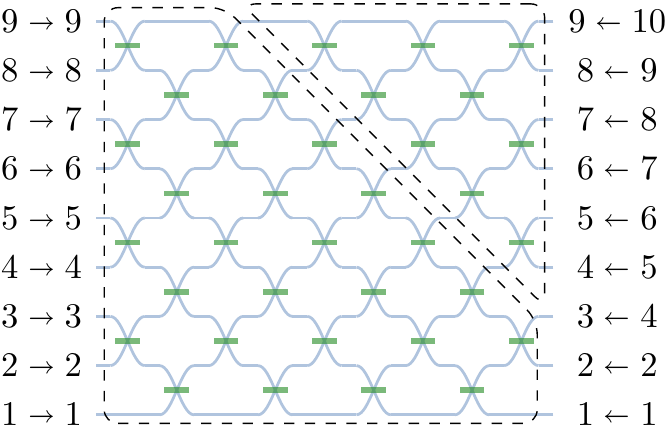}}\\
\subfloat{\includegraphics[scale = 0.85, valign = c]{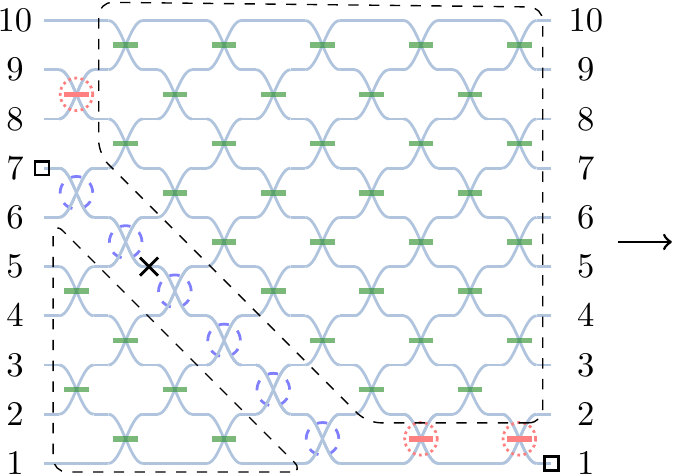}}\hspace{1mm}
\subfloat{\includegraphics[scale = 0.85, valign = c]{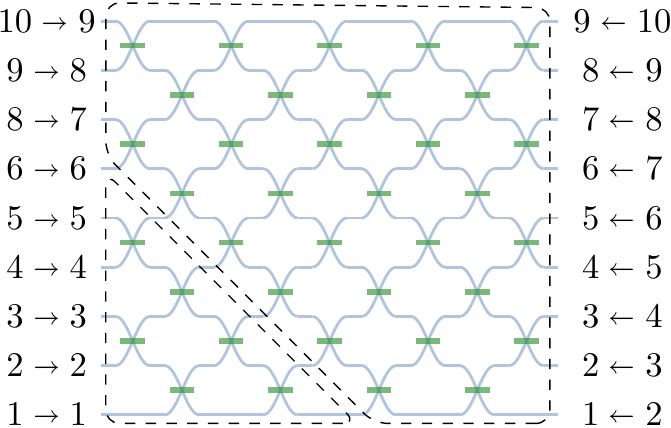}}\\
\subfloat{\includegraphics[scale = 0.85, valign = c]{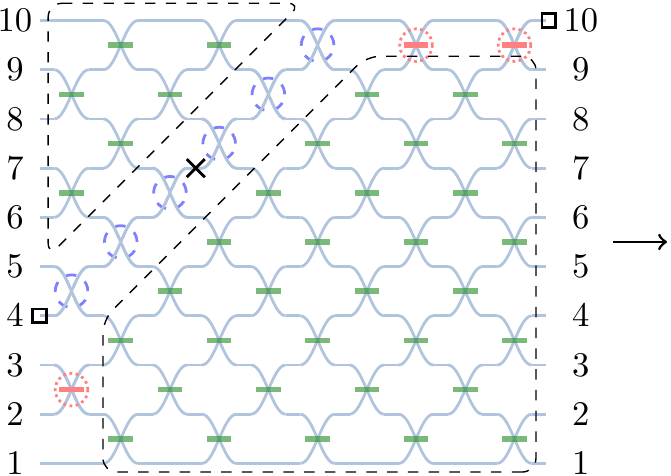}}\hspace{1mm}
\subfloat{\includegraphics[scale = 0.85, valign = c]{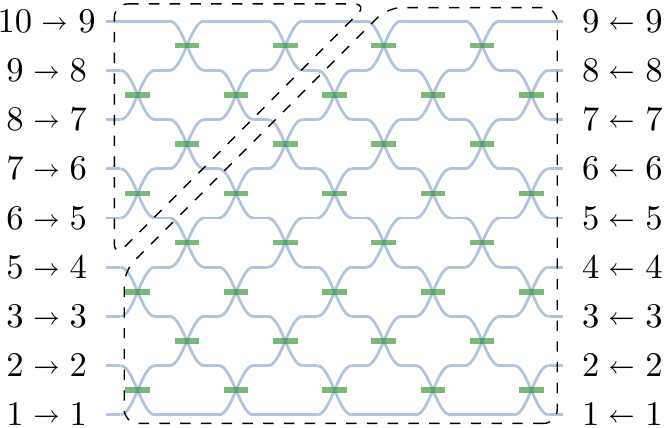}}\\
\subfloat{\includegraphics[scale = 0.85, valign = c]{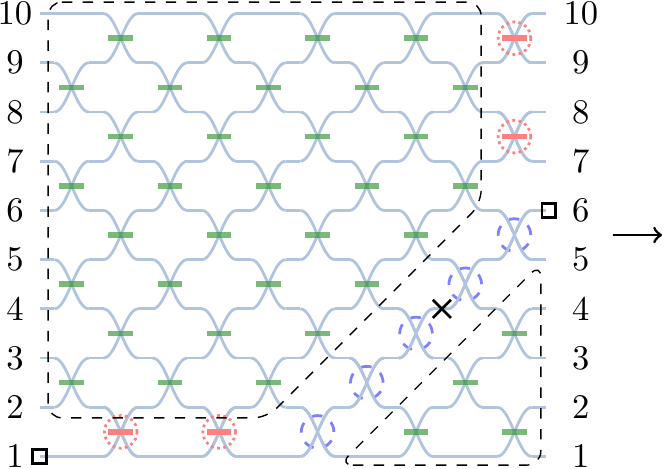}}\hspace{1mm}
\subfloat{\includegraphics[scale = 0.85, valign = c]{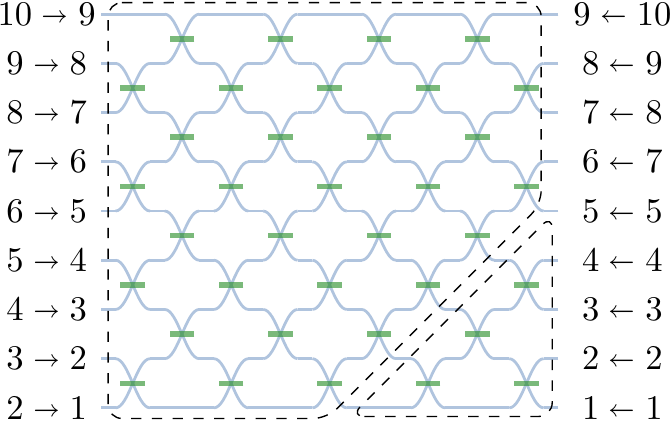}}
\caption{\label{Fig:SingleModeDiag}
\textbf{Method for circumventing a single-mode defect.}
Linear optical interferometers are depicted as a network of beam-splitters (green) acting on multiple modes (grey).
The left column depicts the original interferometers with a cross marking the position of the defect to be circumvented in a ten-mode interferometer.
In the first two rows, this defect is assumed to be in a Northwest--Southeast path, i.e.,~on a mode between the right output port of one beam-splitter and the left input port of a subsequent beam-splitter. 
The last two rows deal with defects in a Northeast--Southwest path, which are between the right output port of one beam-splitter and the left input port of a subsequent beam-splitter.
The odd and even rows depict defects that are situated above and below the main diagonal respectively.
To circumvent a defect, firstly the beam-splitters that act on a diagonal path containing the defect are tuned to fully transmissive (blue dashed circles with no beam-splitter inside).
Additionally, some other beam-splitters are tuned to be fully reflective (red beam-splitters inside red dotted circles) as detailed in the main text.
One input port and one output port (marked with squares) are left unused in the interferometer.
Because of the pattern of fully transmissive and fully reflective beam-splitters, any light that impinges at the unmarked input ports does not interact with the defect and is emitted from only the unmarked output ports.
After discarding the marked input and output ports, the remaining network acts as a smaller linear interferometer that is depicted in the right column.
The input and output ports are relabeled as $m \shortrightarrow n$ and $n \shortleftarrow m$ respectively, where $m$ is the original port label and $n$ is the new port label for the smaller effective interferometer.
}
\end{figure*}

The method to circumvent a defect acting only on one mode is depicted in \cref{Fig:SingleModeDiag}.
Before presenting the details, let us note that discussion of this section is in the context of the rectangular decomposition due to Clements \textit{et al.}~\cite{Clements2016a} in our example but the method can be used for triangular decompositions of Reck~\textit{et al.}~\cite{Reck1994a} or de Guise~\textit{et al.}~\cite{Guise2018} as well.
Recall that these decompositions provide a method for realizing arbitrary $\text{SU}(n)$ transformations on the state of light in an $n$-mode interferometer using $n(n-1)/2$ beam-splitters and $n(n-1)/2 + n-1$ phase-shifters.
For clarity of presentation, only beam-splitters are depicted in \cref{Fig:SingleModeDiag}.

Suppose we are given an interferometer with one defect acting on a single mode at a given location.
In practice, the location of such a defect can be inferred via a calibration procedure such the one presented in~\cite{[{See Section~4 of Supplemental Material for}][~]Harris2017} or via a full tomography procedure such as those presented in Refs.~\cite{Laing2012,Rahimi-Keshari2013,Dhand2016}.

\begin{figure*}
\centering
\subfloat{\includegraphics[scale = 0.84, valign = c]{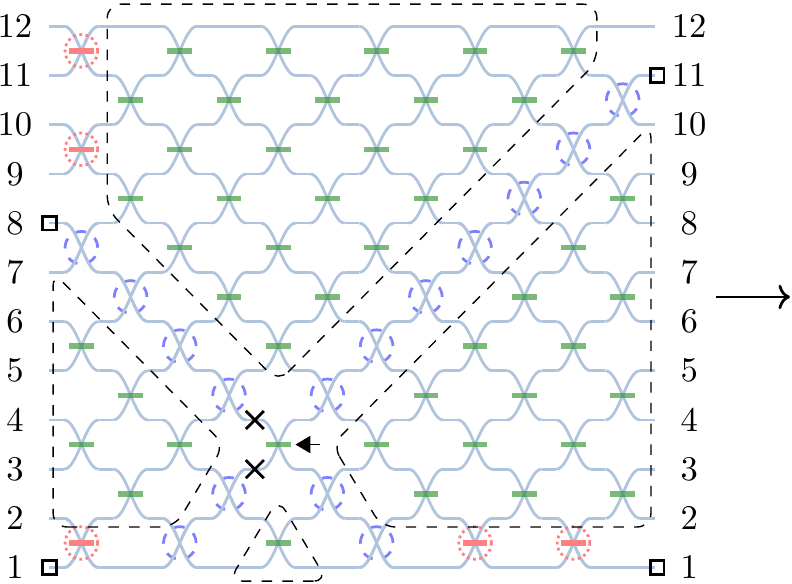}}~
\subfloat{\includegraphics[scale = 0.84, valign = c]{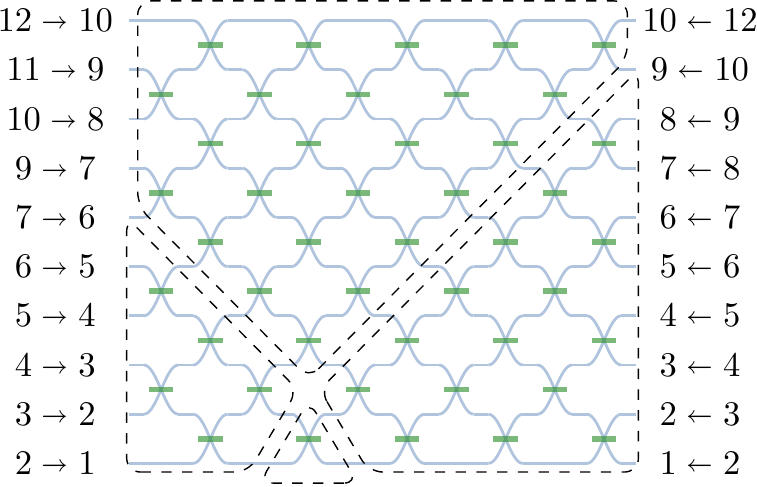}}\\
\caption{\label{Fig:MZI}
\textbf{Method for circumventing defective MZI}
In the left panel, a defective MZI (marked with a triangular arrow pointing towards it) can be circumvented by considering it as two single-mode defects (marked as crosses) and treating these two independently using the method of \cref{Sec:Single}.
After fixing some of beam-splitters as transmissive and reflective, and discarding some of the input and output ports (using similar notation as that in \cref{Fig:SingleModeDiag}), a smaller effective interferometer is obtained as depicted in the right panel.
}
\end{figure*}

The central idea behind the method is to prevent any light from interacting with the defect.
To accomplish this, one input port and one output port are discarded in such a way that
all the light impinging at this input port would reach the defect, and any light that would pass the defect (i.e., the defective mode at the point of the defect) would be emitted from the chosen output port.
In other words, there is a direct path for all the light impinging at the chosen input ports to be emitted from the chosen output port and, importantly, that the defect lies on that path.
Such a path can be obtained by setting some specific beam-splitters to be perfectly transmissive while others are set to be perfectly reflective as described in the remainder of this section.
Once this path and ports are chosen, then light is impinging only at the remaining input ports and is detected at the output of the remaining output ports.

Now we are ready to discuss a detailed prescription for choosing the input and output ports and connecting them with a direct path through the defect.
As depicted in the left column of \cref{Fig:SingleModeDiag}, the beam-splitters in the diagonal path containing the defect are made fully transmissive, i.e., all the light from one mode of the beam-splitter is transmitted into the other mode.
Some other beam-splitters are set to  full reflectivity depending on one of the two cases as follows.
Note that this discussion uses the convention that the light in the interferometer runs from left to right.

The first case is that of a defect in a Northwest--Southeast diagonal path, i.e.,~on a mode between the lower output port of one beam-splitter and the upper input port of a subsequent beam-splitter. 
In this case, let us focus on the points where the diagonal path meets the edges of the interferometer, where interferometer edges refer to (a.)~the beam-splitters acting on the first 
or the last mode at the top or the bottom of the circuit; (b.)~those that are just to the right of the input ports; and (c.)~those that are just before the output ports.
Each of the beam-splitters between the top left end of the path and the beam-splitter at the top-left corner of the interferometer (i.e., the first beam-splitter acting on the last mode) are set to be fully reflective.
Similarly, each of the beam-splitters between the bottom right point of the path and the beam-splitter at the bottom-right corner of the interferometer (i.e., the last beam-splitter acting on the first mode) are set to be fully reflective.
The phase-shifters along the path can be set to any value.
This situation is depicted in the first two rows of \cref{Fig:SingleModeDiag}.

Now consider the second case, that of a defect in a Northeast--Southwest diagonal path, i.e., on a mode between the right output port of one beam-splitter and the left input port of a subsequent beam-splitter.
Here as well, the beam-splitters between the points where the diagonal path meets the edges and the bottom-left and top-right beam-splitters are made fully reflective.
As before, the phase-shifters along the chosen path can be set to any value.
This case is depicted in the bottom two rows of \cref{Fig:SingleModeDiag}.

Because of this pattern of fully reflective and transmissive beam-splitters, one input port and one output port (marked with squares in the figure) are connected to each other via a direct path.
Any light impinging at the marked input port would transverse the path and pass through the defect and finally arrive only at the marked output port.
Because of the unitarity of the transformation, this also means that any light impinging at the other input ports would not traverse the path and, crucially, would not interact with the defect.
Finally, this light would only emitted only from the other output ports.

The remaining beam-splitters and phase-shifters can be used to implement a universal $\text{SU}(n-1)$ transformation.
First, let us count the number of remaining beam-splitters and phase-shifters. 
As $n-1$ beam-splitters are set to a fixed value in the procedure, these can no longer modified based on the transformations that are desired to be implemented.
Thus, $n(n-1)/2 - (n-1) = (n-1)(n-2)/2$ tunable beam-splitters remain free, which is exactly the number required for implementing an $(n-1)$-mode universal transformation.
Similarly, as $n$ phase-shifters can no longer be used for implementing arbitrary transformations, the remaining number of freely-tunable phase-shifters is $(n-1)(n-2)/2 + (n-2)$, which is just the right number required for implementing $(n-1)$-mode transformations. 
Furthermore, as can be seem by inspection of \cref{Fig:SingleModeDiag}, we see that the components are are still connected together in the correct structure, i.e.,~identically to the structure of a universal interferometer that acts on one less mode than the original interferometer.

\begin{figure}
\centering
\subfloat{\includegraphics[scale = 0.94, valign = c]{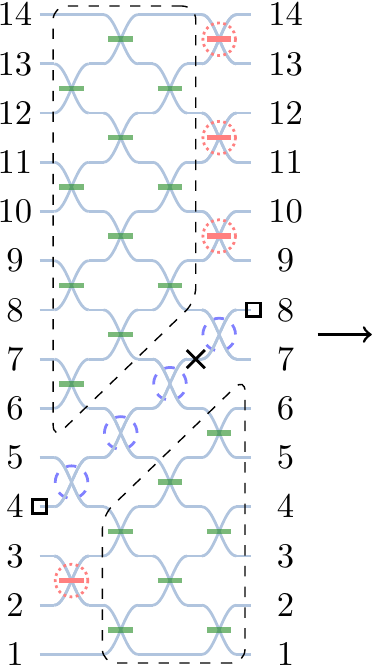}}~
\subfloat{\includegraphics[scale = 0.94, valign = c]{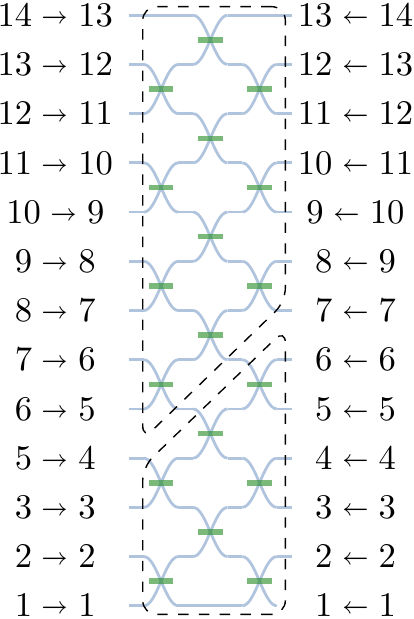}}\\
\caption{\label{Fig:shallow}
\textbf{Method for circumventing single-mode defect in shallow circuits.}
That notation is the same as that of \cref{Fig:SingleModeDiag}.
In the left panel, a 14-mode depth-4 circuit has a single-mode defect.
This defect can be circumvented to obtain a 13-mode depth-3 circuit, as shown on the right.}
\end{figure}

Hence, we see that any smaller $(n-1)$-mode transformations can be realized using the larger $n$-mode interferometer with a single defect.
Specifically, given an $(n-1)$-mode transformation that is to be implemented, as a first step, the beam-splitter and phase-shifter parameters to realize this are determined using a decomposition, for example the decomposition due to Clements \textit{et al.}
Next, the specific beam-splitters on the chosen path are respectively set to be fully reflective and fully transmissive as described above, while the remaining are set according to the parameters obtained for the $(n-1)$ mode transformation.
The input and output modes are relabeled as depicted in the right column of \cref{Fig:SingleModeDiag},
With this relabeling, the $(n-1)$ new input ports are connected to the new $(n-1)$ output ports via the desired $(n-1)$ mode transformation acting between the ports.

\subsection{Circumventing multiple defects}
\label{Sec:Multi}
The method for circumventing a single defect can be used repeatedly to also circumvent multiple defects, each acting on one or more modes.
In fact, it turns out that the only case to be addressed is that of multiple defects such that each acts on a single mode.
This is because those situations in which more than one mode is defective (for example the case of a defective beam-splitter) are completely equivalent to multiple single-mode defects that act on the modes leading to the multi-mode defect.
As an example, consider \cref{Fig:MZI} which shows a defective beam-splitter, marked with an arrow.
Circumventing this (two-mode) beam-splitter is equivalent to circumventing two single-mode defects that are present at the two input ports of the beam-splitter.
If no light enters these virtual single-mode defects, then no light will enter the two-mode beam-splitter.
Thus, we can focus on the problem of circumventing one or more single-mode defects because any situation of multiple single- or multi-mode defects can be reduced to such a problem.

For each of the single-mode defects, the procedure described above is applied independently.
I.e., their respective diagonal paths are followed, and each of the beam-splitters along these paths are set to be fully transmissive.
Once the diagonals reach the edge, follow the edge to the top-left and bottom-right beam-splitters (bottom-left and top-right) for a diagonal path in the Northwest-Southeast (Northeast-Southwest) direction, setting all beam-splitters along the way to perfectly reflective.
If the instructions from two or more defects are conflicting, for example, if one defect requires setting the beam-splitter to be reflective and another defect requires it to be transmissive, then any setting of the beam-splitter can be chosen as no light will interact with this beam-splitter.

For each single-mode defect, the method will discard one input and one output port.
Thus, an $n$-mode universal interferometer with $m$ single-mode defects can be used to implement universal transformations on $n-m$ modes.
An example of this is an original $n=12$ mode interferometer with a defective beam-splitter, which corresponds to $m=2$ single-mode defects.
Such an interferometer can be used to implement an effective $12-2=10$ mode interferometer as illustrated in \cref{Fig:MZI}.

\subsection{Defects in shallow circuits}
\label{Sec:Shallow}

The circuits considered thus far have corresponded to universal interferometers, i.e., ones that can be used to implement arbitrary linear optical unitary transformations.
These circuits act on as many modes as their circuit depth, which is defined by the maximum number of beam-splitters that may act on light as it transverses the circuit.
Other types of circuit geometries may also be considered, particularly those with lower circuit depth than the number of modes.
These so-called ``shallow'' circuits are relevant for generating tensor-network states of light~\cite{Dhand2018,Lubasch2018}.
Such shallow circuits can be implemented in a modular manner as proposed by Mennea \textit{et al.} in~\cite{Mennea2018}.

Our procedure can also be exploited to circumvent defects in such shallow circuits.
An example of such a defect and its circumvention is depicted in \cref{Fig:shallow}.
As in the case of deep ($n$-mode, $n$-depth) circuits, circumventing single-mode defects decreases the number of modes by one.
Moreover, the circuit depth decreases by one as well.
As before, $m$-mode defects, such as defects in beam-splitters, can be circumvented at the cost of reducing both the number of modes and circuit depth by $m$.

\section{Improvement of defect tolerance}
\label{Sec:Disc}

\begin{figure}
\includegraphics[width = \columnwidth]{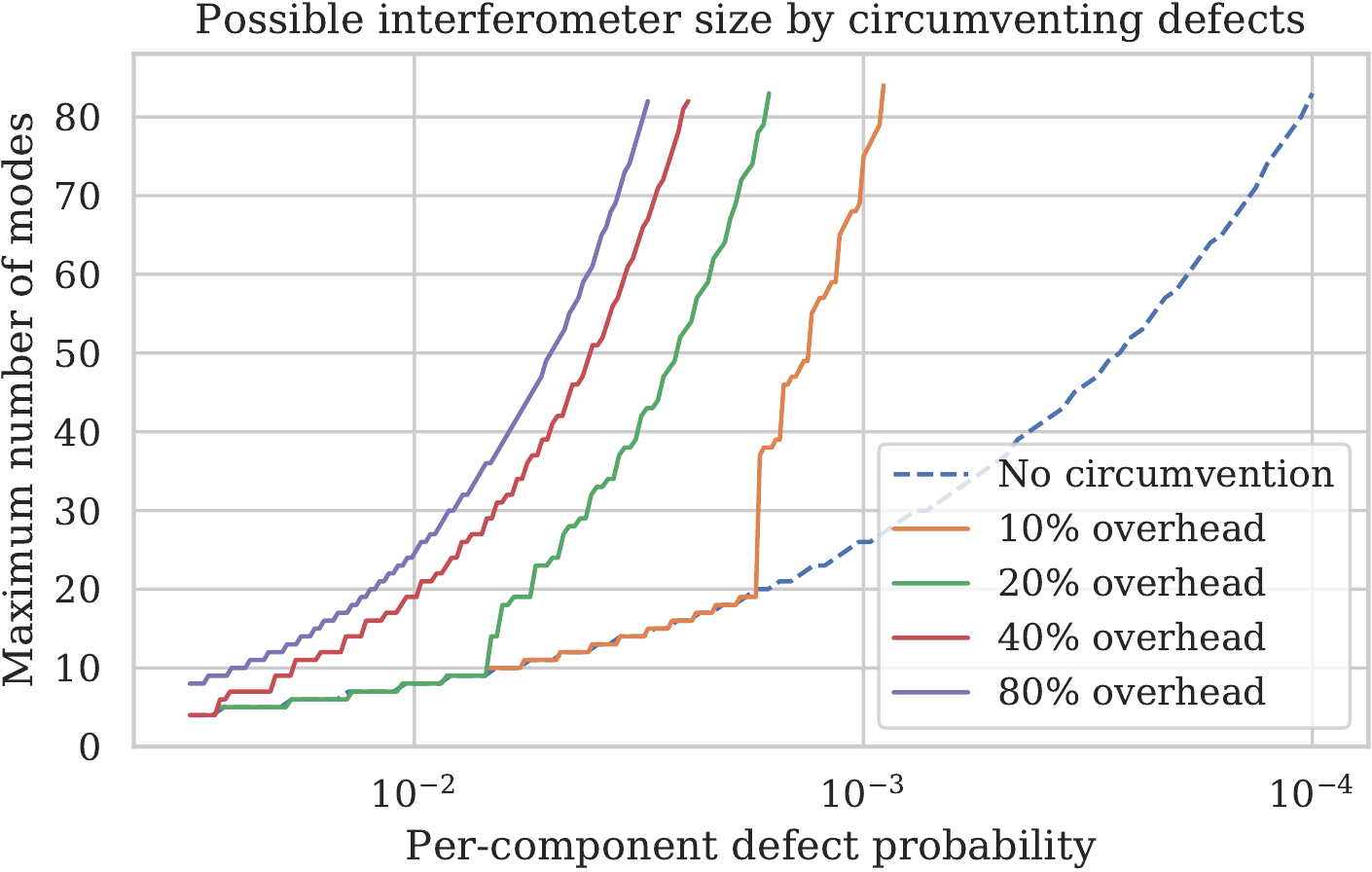}

\caption{Increase in defect-tolerance as a result of the circumvention procedure for different values of overhead, which is quantified by the ratio $m/n$.
The size of the largest universal unitary transformation that is possible to construct (with at least $50\%$ success probability) is plotted as a function of the per-component defect probability.
The blue dashed curve represents the scaling if no circumvention is used.
The remaining curves are obtained by constructing a larger $(n+m)$-mode interferometer, and allowing for up to $m$ defects.
These defects are circumvented, thus resulting in an $n$-mode universal interferometer. 
The percentage value in the legend is the allowed overhead quantified by $m/n$.
\label{Fig:defects}}
\end{figure}

By allowing the use of interferometers with one or more defects, the circumvention procedure drastically increases the acceptable probability of defects in components.
To put it another way, for a given defect probability, the circumvention procedure enables constructing interferometers acting universally on significantly larger numbers of modes than would otherwise be possible.
Suppose that an application requires a universal interferometer with $n$ modes.
Let us compare the situation in which no circumvention procedure is used with one in which we are allowed to circumvent defective components.

The first case is one in which no circumvention procedure is available, and even a single defect in the interferometer precludes its use. 
What is the maximum allowed per-component defect probability $\epsilon$ such that more than $50\%$ of the constructed $n$-mode interferometer have zero defective optical components?
In other words, what is the per-component defect tolerance such that a majority of the implemented interferometers are functional?
The number of optical components on the interferometer is $\approx n^{2}$ including phase-shifters and beam-splitters.
Thus, the probability that none of the components has any defect is 
\begin{equation}
	(1-\epsilon)^{n^{2}} \stackrel{!}{>} 1/2,
\end{equation}
which implies the following condition on maximum number of modes that can be implemented is
\begin{equation}
	n < \sqrt{\frac{\log(2)}{\log(1/(1-\epsilon))}}.
\end{equation}
Note that this is a severe requirement and requires defect rates smaller than one in $10^{3}$ for implementing interferometers with more than 25 modes or so.
This requirement is presented graphically in the blue dashed curve of \cref{Fig:defects}

Now consider a situation were the defects can be circumvented using the current procedure.
To implement an $n$-mode transformation, the actual interferometer is constructed with $(n+m)$ modes, where the extra $m$-modes improve the defect tolerance.
This extra number of modes describes the \textit{overhead} of our procedure, as defined by the ratio $m/n$.
After the circumvention procedure, each defect reduces the number of modes by one or two, so up to $m$ single-mode or $\lfloor m/2 \rfloor$ two-mode defects can be tolerated.
Here, the total number of components to be implemented increases to $\approx (n+m)^{2}$.
For concreteness, let us focus on single-mode defects.
In this case, the relevant probability required to be more than $50\%$ is that of having $m$ defects or fewer, which is 
\begin{align}
\begin{split}
	&(1-\epsilon)^{(n+m)^{2}} + (n+m)^{2}\epsilon(1-\epsilon)^{(n+m)^{2}-1} + \dots \\
	&\quad + \frac{(n+m)^{2}!}{((n+m)^{2}-m)!m!}\epsilon^{m}(1-\epsilon)^{(n+m)^{2}-m} \stackrel{!}{>} 1/2.
\end{split}
\end{align}
Depending on the chosen values of $m$ and $n$, this is a far less stringent requirement, as plotted in \cref{Fig:defects}.
Note that by allowing an overhead $m/n$ of between $10$\% and $80$\%, the tolerable defect threshold can increase by more than an order of magnitude.
Considering a fixed defect probability, we see a significant enhancement in the maximum size of interferometer that can be constructed. 

\section{Summary}
\label{Sec:Summary}
In summary, this work introduces a procedure for the circumvention of defective components in a linear optical interferometers.
The procedure allows for exploiting an $(n+m)$-mode universal but defective interferometer with $m$ defects as an effective $n$-mode universal interferometer.

As the procedure prevents any light from interacting with the defect, it is agnostic to the type of defect being mitigated.
In particular, it could be used for circumventing the most common defect types, including not only lossy modes but also phase-shifters and beam-splitters with imperfect reconfigurability. 
Furthermore, the procedure could also find applicability in the field of multi-port optical switches being developed for use in data centers~\cite{Cheng2018}.
The building blocks of such switches include not just MZIs but also switches based on micro-electromechanical system or on micro-ring resonators and our procedure could be exploited to circumvent defects in such components as well.
Another potential use case of the procedure is towards optical neural networks, which rely on the action of linear optical interferometers on classical or quantum light~\cite{Shen2017,Tait2017,Steinbrecher2019}.

One approach to exploiting the procedure could be to implement a smaller unitary matrix by fabricating larger interferometers and circumventing the defects the arise therein.
In this approach, the procedure enables a remarkable improvement in the maximum tolerable defect probability in the construction of large interferometers.

\acknowledgements{
I thank Kamil Br\'{a}dler, Dylan Mahler, Blair Morrison, Kang Tan, and Zachary Vernon for helpful discussion and for bringing this problem to my attention. Thanks to Lukas G.~Helt, Yanbing (Young) Zhang and David Asgeirsson for a careful proof-reading of the manuscript and insightful comments. 
}{}
%

\end{document}